\documentclass[fleqn,usenatbib,letters]{mnras}

\usepackage{newtxtext,newtxmath}
\usepackage[T1]{fontenc}
\usepackage{ae,aecompl}
\usepackage{longtable}
\usepackage{graphicx}	% Including figure files
\usepackage{amsmath}	% Advanced maths commands
\usepackage{amssymb}	% Extra maths symbols
\usepackage[utf8]{inputenc}

\newcommand{\Teff}   {$T_{\rm eff}$~}

\newcommand{\logg}{$\log g$} 
\newcommand{\feh}   {$[\mathrm{Fe/H}]$~}

\newcommand{\Prot}   {$P_{\rm rot}$~}
\newcommand{\Psini}   {$P_{\rm rot}/\sin i$~}
\newcommand{\vsini}   {$v\sin i$~}
\newcommand{\kepler} {\textit{Kepler}}
\newcommand{\gaia} {\textit{Gaia}}

\title[]{Constraining the evolution of stellar rotation using solar twins}
\author[Evolution of stellar rotation using solar twins]{Diego Lorenzo-Oliveira,$^{1}$\thanks{E-mail: diegolorenzo@usp.br}
Jorge Meléndez,$^{1}$
Jhon Yana Galarza,$^{1}$
Geisa Ponte$^{2}$,\newauthor 
Leonardo A. dos Santos$^{3}$, Lorenzo Spina$^{4}$,
Megan Bedell$^{5}$,
Iv\'an Ram\'{i}rez$^{6}$, \newauthor Jacob L. Bean$^{7}$ and
Martin Asplund$^{8}$
\\
$^{1}$Universidade de S\~ao Paulo, Departamento de Astronomia do IAG/USP, Rua do Mat\~ao 1226, Cidade Universit\'aria, 05508-900 \\S\~ao Paulo, \ SP, Brazil \\
$^{2}$Universidade Federal do Rio de Janeiro, Observat\'orio do Valongo, Ladeira do Pedro Antonio 43, CEP: 20080-090 Rio de Janeiro,\\
RJ, Brazil\\
$^3$Observatoire astronomique de l’Université de Genève, 51 chemin des Maillettes, 1290 Versoix, Switzerland \\
$^4$Monash Centre for Astrophysics, School of Physics and Astronomy, Monash University, VIC 3800, Australia\\
$^5$Center for Computational Astrophysics, Flatiron Institute, 162 5th Ave., New York, NY 10010, USA\\
$^6$Tacoma Community College, 6501 South 19th Street, Tacoma, Washington 98466, USA\\
$^7$University of Chicago, Department of Astronomy and Astrophysics, USA\\
$^8$The Australian National University, Research School of Astronomy and Astrophysics, Cotter Road, Weston, ACT 2611, Australia
}

\date{Accepted XXX. Received YYY; in original form ZZZ}

\pubyear{2019}

\begin{document}
\label{firstpage}
\pagerange{\pageref{firstpage}--\pageref{lastpage}}
\maketitle

\begin{abstract}
The stellar Rotation $vs.$ Age relation is commonly considered as a useful tool to derive reliable ages for Sun-like stars. However, in the light of \kepler\ data, the presence of apparently old and fast rotators that do not obey the usual gyrochronology relations led to the hypothesis of weakened magnetic breaking in some stars. In this letter, we constrain the solar rotation evolutionary track  using solar twins. Predicted rotational periods as a function of mass, age, \feh and given critical Rossby number ($Ro_{\rm crit}$) were estimated for the entire rotational sample. Our analysis favors the smooth rotational evolution scenario and suggests that, if the magnetic weakened breaking scenario takes place at all, it should arise after $Ro_{\rm crit}\gtrsim2.29$ or ages $\gtrsim$5.3 Gyr (at 95\% confidence level).
\end{abstract}
%We compare the smooth rotational evolution that anchors the gyrochronology relations with the weakened magnetic breaking scenario. Comparing both approaches with our solar twin data, we tend to favor the smooth solar rotational evolution, albeit with a marginal significance. On the other hand, our results also suggest that, if the magnetic weakened breaking scenario is taking place at a given evolutionary stage, it should occur after $Ro_{\rm crit}\sim2.2$ (at 95\% confidence) which corresponds to an age slightly older than the Sun.
\begin{keywords}
Sun: rotation – stars: solar-type – stars: rotation – stars: fundamental parameters
\end{keywords}

%%%%%%%%%%%%%%%%%%%%%%%%%%%%%%%%%%%%%%%%%%%%%%%%%%

%%%%%%%%%%%%%%%%% BODY OF PAPER %%%%%%%%%%%%%%%%%%

\section{Introduction}
Rotation-based ages of old Sun-like stars are rooted in a complex and intricate dependence on age, rotation, turbulent convection, structural variations and mass-loss due to magnetized winds \citep{skumanich72,reiners12,guerrero13, fionnagain18}. Classically, the age-dating method that relies on this phenomenon assumes that the rotational periods ($P_{\rm rot}$) can be expressed in well-defined functions of the age and mass (or a proxy of it), the so-called gyrochronology relations \citep{barnes07,mamajek08}. These relations had successfully confirmed the paradigm of rotation-activity-age coupling that powers the global dynamo evolution along the main-sequence \citep{barnes07,vidotto14,donascimento14,lorenzo16,lorenzo18} and reproduced the main features observed in open clusters spanning a wide range of ages \citep{meibom15}.

Apart from this inspiring agreement, some of the old \kepler\ field stars shows unexpected fast rotation, especially hotter ones with ages greater than 2$-$3 Gyr \citep{angus15,metcalfe16}. This tension led to idea that after a critical Rossby number \citep[$Ro \equiv P_{\rm rot}/\tau_{\rm CZ}$, where $\tau_{\rm CZ}$ is the convective turnover time;][]{noyes84} a drastic change of the stellar differential rotation (SDR) pattern might hamper the production and maintenance of magnetic field large-scale components over secular timescales. One of the most important (and accessible) effects  of this drastic transition would be the presence of old and fast rotating stars with reduced angular momentum loss caused by magnetized winds. However, recent observational results gave us alternative hints about the possible smooth nature of the Sun-like rotational evolution in the light of \kepler\ asteroseismic data \citep{benomar18}.

Motivated by this up-to-date discussion about the smooth nature of the age-rotation relations, we can ask ourselves: \textit{what solar twins can tell us about the solar rotational evolution}? This letter uses solar twins to evaluate a recent claim \citep{vansaders16} about the solar rotational transition at a given critical Rossby number. Sec. \ref{sec:rotation_sample} describes our working sample, selection criteria of solar twin rotators and determination of \Prot through activity time series. In Sec. \ref{sec:age_rotation} we discuss the age-rotation evolution of solar twins and the suitability of standard rotational evolution models. The conclusions are drawn in Sec. \ref{sec:conclusions}.

\section{Solar twins rotation sample}\label{sec:rotation_sample}

We compiled 79 solar twins (plus the Sun) presented in \citet{spina17}. Our sample was extensively observed over the years (2011$-$2016) with HARPS spectrograph \citep{mayor03} fed by the 3.6 m telescope at La Silla Observatory, to search  for planets around solar twins \citep[program 188.C-0265,][]{melendez17}. To prevent the inclusion of anomalously fast rotators due to binarity effects, 17 spectroscopic binary (SB) stars and other solar twins with a close companion within 4'' were discarded from the analysis \citep{santos17}. Other interesting solar twins spectroscopically analyzed in the past by our group were added to our sample \citep[HIP30503 and HIP78399 in][]{galarza16}. The resulting HARPS sample of this work is composed of 65 stars. 

In order to estimate rotational velocities and other stellar parameters of interest, we made use of the updated atmospheric parameters ($T_{\rm eff}$, \logg, $[\mathrm{Fe/H}]$, and $[\mathrm{\alpha/Fe}]$) provided by \citet{spina17}. \gaia\ DR2 \textit{G} band photometry and parallaxes \citep{gaia18} were combined to the spectroscopic data to obtain stellar ages, masses, \logg\ and other evolutionary parameters following the procedures described in \citet{nolan18} (see their Sec. 6). For distant \kepler\ solar twins that will be described in the following sections, we used the reddening \textit{G} band corrections provided by \gaia\ DR2. The typical mass and metallicity of our solar twin sample agrees with solar parameters within $\pm$0.05 $M_{\odot}$ and $\pm$0.04 dex, respectively. Stellar atmospheric and evolutionary parameters, projected \Prot and other relevant information of the entire sample is shown in Table \ref{tab:data}. Projected rotational and macroturbulence velocities were determined through HARPS \textit{full width half maximum} of the \textit{cross correlation function} ($FWHM_{\rm CCF}$) vs. \vsini together with macroturbulence calibrations provided by \citet{santos16}. For details about the adopted procedures, see their Sec. 3. The CCF based \vsini errors are computed propagating the $FWHM_{\rm CCF}$ calibration and typical \vsini measurement errors estimated by \citet{santos16} yielding $\sigma$($v\sin i$)=0.23 km/s or $\sigma$($P_{\rm rot}/\sin i$) of 3 days (assuming 2\% of $\sigma (R/R_\odot)$).
\begin{table*}
	\caption{Relevant parameters for our rotation sample of solar twins stars analyzed in this paper. This table is available in its entirety in machine readable
		format at the CDS.}
	\begin{tabular}{lcccccccl}
		\hline
		HIP & Age & Mass & \feh & \Psini  & \Prot &  $P_{\rm rot}^P$  & $P_{\rm rot}^P (Ro_{\rm crit}=2.0)$ & Remark \\
		-- & [Gyr] & [$M/M_\odot$] & -- & [d] & [d] & [d] & [d] &-- \\
		\hline
		\hline
1954 &   4.3$_{-0.3}^{+0.3}$   &   0.97$\pm$0.03   &   -0.090$\pm$0.003   &   26.6$\pm$3.1   &   24.1$\pm$0.2   & 25.2$\pm$2.7 & 24.8$\pm$2.7 &   This work \\ 
7585 &   4.1$_{-0.2}^{+0.3}$   &   1.03$\pm$0.03   &   0.083$\pm$0.003   &   24.1$\pm$3.0   &   23.0$\pm$2.3   & 23.4$\pm$3.4 & 20.3$\pm$3.9 &   \citet{see17} \\ 
22263 &   0.6$_{-0.6}^{+0.5}$   &   1.06$\pm$0.03   &   0.037$\pm$0.006   &   14.8$\pm$3.0   &   11.8$\pm$3.0   & 9.6$\pm$4.3 & 9.6$\pm$4.3 &   \citet{suarez17} \\ 
30503 &   3.0$_{-1.4}^{+0.7}$   &   1.08$\pm$0.04   &   0.070$\pm$0.016   &   27.3$\pm$3.0   &   20.0$\pm$0.1   & 18.7$\pm$4 & 15.6$\pm$3.8 &   This work \\ 
36515 &   0.3$_{-0.3}^{+0.3}$   &   1.03$\pm$0.03   &   -0.029$\pm$0.009   &   11.9$\pm$3.0   &   4.6$\pm$1.2   & 7.2$\pm$2.3 & 7.2$\pm$2.3 &   This work \\ 
	
	\hline			
	\label{tab:data}
\end{tabular}
\end{table*}

Given that the rotation axes of the stars are randomly oriented in different inclination angles ($i$), our spectroscopic analysis is limited by the projection factor $\sin i$ \citep{gray05}. This feature skews the \Prot distribution towards higher values of rotation for a given age owing to the geometric factor 1/$\sin i$. On the other hand, if one analyses a progressively larger sample of stars, it is indeed expected that, for a given age/mass/${\rm [Fe/H]}$, the lower boundary of \Psini distribution asymptotically matches with the \textit{true} distribution of \Prot which is scattered by intrinsic effects (e.g. propagation of initial conditions and SDR effects). In fact, this approximation has an optimal applicability for samples of twin stars like ours, where mass and \feh effects, that might hamper the statistical corrections towards the true \Prot distributions, are mitigated. 

\begin{figure}
	\centering
	\begin{minipage}[t]{0.45 \linewidth}
		\centering
		\resizebox{\hsize}{!}{\includegraphics{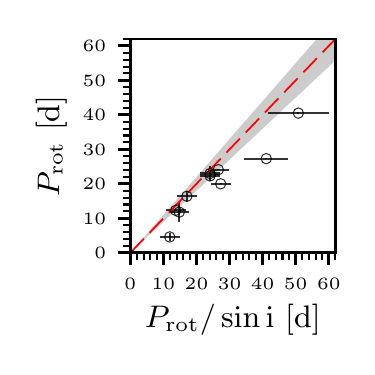}}
	\end{minipage}
\centering
\begin{minipage}[t]{0.45 \linewidth}
	\centering
	\resizebox{\hsize}{!}{\includegraphics{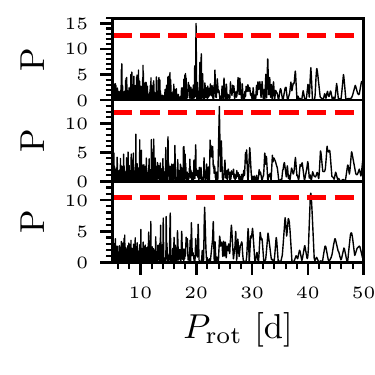}}
\end{minipage}
	
	\caption{\textit{Left panel}: Comparison between \Prot and \Psini for a subsample of 10 solar twins observed by HARPS. The dashed red line stands for 1:1 relation and shaded region represents 10\% of \Prot intrinsic errors. \textit{Right panels}: \ion{Ca}{II} periodogram analysis of three solar twins with different ages. The most probable peaks for HIP30503 (upper panel), HIP1954 (middle panel), and HIP118115 (lower panel) are 20.0, 24.1, and 40.5 days. The dashed red lines stand for the false-alarm probability of 3 sigma.}
	\label{fig:psini_prot_fig}
\end{figure}

Through 10$^6$ Monte-Carlo (MC) simulations we sought for an optimal selection criterion that closely unveils the distribution of stars with the highest chance of having ${\rm \sin i}\sim$1, for a given sub-sample size $N$. We ran $N$ MC simulations assuming random angle orientations ranging from 0 to $\pi$/2 for a range of angular rotational frequencies ($\Omega$). We assigned an intrinsic $\Omega$ error of 10\% to account for stellar differential rotation \citep[SDR,][]{epstein14}. For each simulation, the difference ($\Delta$) between the \textit{true} \Prot and the median \Psini distribution delimited by its upper 0.5(P50), 0.7(P70), 0.84(P84), and 0.975(P98) percentiles cut-offs. This procedure was repeated 10$^4$ times in order to estimate the best selection criteria (using the sample medians of each upper percentiles cut-offs) that minimizes $|\Delta|$ $-$ i.e. the stellar inclination selection bias. For a typical sub-sample size of 10 stars, we found that the average between P70 and P98 estimates closely converges into the centroid of the \textit{true} \Prot distribution, assuming rotation roughly constant along the age domain considered. The exclusion of the upper 2.5\% of each simulation prevents to systematically include unusually fast-rotating stars in each age-bin. Following this selection prescription, we bootstrapped the 1$\sigma$ confidence intervals within each age-bin of 2 Gyr through 10$^4$ repetitions. We found 2 Gyr age-bin as our optimal choice because it balances sampling and the ratio between the expected \Prot evolution and realistic \Prot errors due to intrinsic effects, at least for stars older than $\sim$1 Gyr. In fact, during the simulations, we also compute the age-\Prot correlation within each age bin and found negligible correlation between both variables (p-value$\gg$0.05). In Table \ref{tab:data}, we highlight the stars with probable $\sin~i\sim1$ along each age-bin. In Fig. \ref{fig:psini_prot_fig} (left panel) we show our subsample of 10 stars with both HARPS \Psini and \Prot measurements. Our \Psini estimates are consistent with the 1:1 relation represented by the red dashed lines. The intrinsic errors associated to \Prot measurements (10\%) are given by shaded region along the 1:1 identity line. As the inclination factor is encapsulated in the \Psini, the projected \Prot are slightly shifted towards higher rotation values albeit it marginally converges into the shaded region indicating that the majority of these stars have $\sin i\sim1$.    

Using our large \ion{Ca}{II} activity time-series of solar twins \citep{lorenzo18}, we determined \Prot of six stars using a Generalized Lomb-Scargle analysis \citep[GLS,][]{zechmeister09}: HIP 1954, 30503, 36515, 79672, 95962 and 118115. Our procedure is similar to the one described by \citet{suarez17}. In brief, for each star, we cleaned our time-series by removing the observations with low signal to noise ratio around the \ion{Ca}{II} lines (SNR$<$30). To avoid the inclusion of poor observations or stellar transient events such as flares, we removed from the activity time-series outliers placed above $\geq2.5\sigma$. Then, we look for the presence of strong signals with bootstrapped false-alarm probability (FAP) $\geq$3$\sigma$ at frequencies related to typical rotational timescales ($\leq$50 days). In other cases where no significant peaks were found, we detrended the time-series from eventual sinusoidal long-term signal (FAP$\geq3\sigma$) that are likely to be associated to stellar cycle modulations. In these cases, the \Prot measurements are determined in subsequent periodogram analysis of the detrended time-series. Signals matching with the expected window function in period space were not considered in our analysis. Figure \ref{fig:psini_prot_fig} (right panels) shows our periodogram analysis for three solar twins. Rotation periods for another four solar twins were gathered from the literature: HIP 7585, 22263, 42333 and 43297 \citep{petit08,wright11,suarez17,see17}.

Additionally, another four solar twins with \kepler\ \Prot were added to our analysis. Precise atmospheric parameters were obtained with Gemini/GRACES (ID:GN-2018B-FT-101) and KECK/HIRES high signal-to-noise ratio and high resolution observations conducted by our group \citep[][Yana Galarza et al. in prep.]{bedell17}: Kepler-11, KIC 10130039, 12404954 and 7202957 \citep{mcquillan13,mazeh15}. The relevant information about the \kepler\ solar twins are summarized in Table \ref{tab:data}. Finally, we collected literature data from three solar metallicity old open clusters observed by the \kepler\ mission spanning a critical age range of rotational evolution: NGC6811 ($\sim$1 Gyr, N=5 stars) NGC6819 ($\sim$2.5 Gyr, N=5 stars) and M67 ($\sim$4 Gyr, N=12 stars) \citep{meibom11, meibom15, barnes16, brandenburg18}. We restricted our sample selection only to those stars with near-solar \Teff based on ${\rm (B-V)}_0$ index \citep[5600 $\leq$ \Teff $\leq$ 5900 K,][]{casagrande10}. For NGC6811, the $\rm{(B-V)}$ colors were estimated inverting the $\rm{(g-r)}$ vs. $\rm{(B-V)}$ calibration equation by \citet{bilir05}. The average \Prot of $\sim$1 solar mass star in these clusters are 10.2$\pm$0.6, 18.1$\pm$0.5 and 24.0$\pm$2.4 days for NGC6811, NGC6819 and M67, respectively.

\section{Age-Rotation relation}\label{sec:age_rotation}
\begin{figure}
\centering
  \begin{minipage}[t]{0.85 \linewidth}
\centering
    \resizebox{\hsize}{!}{\includegraphics{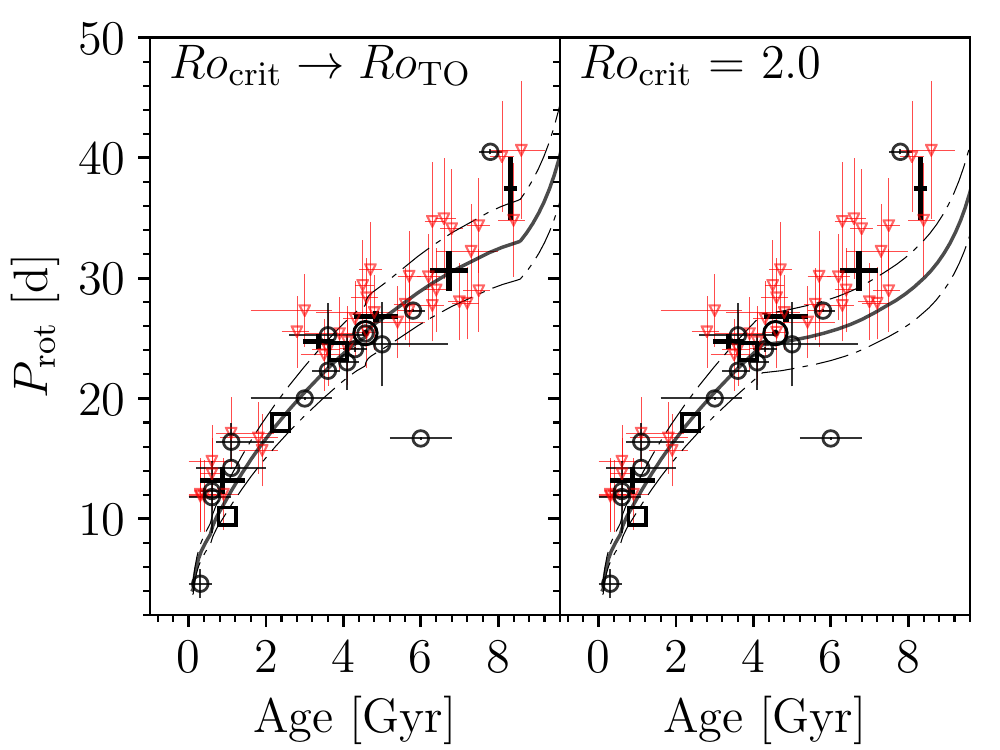}}
  \end{minipage}

  \caption{\textit{Left panel}: Modified Kawaler rotation track for 1.00 M$_\odot$ and solar metallicity in solid black line without considering the weakened braking efficiency scenario. We assign $Ro_{\rm critical}$ at the turn-off as 2.6 ($Ro_{\rm TO}$). Dash dotted lines stand for 10\% of \Prot error given by intrinsic effects. \kepler\ OCs are represented by black squares. Solar twins with measured \Prot are the black empty circles. Selected twin sample and its respective age-\Prot centroid stand for red triangles and black crosses, respectively. The Sun is represented by $\odot$. \textit{Right panel}: Modified Kawaler wind law model for $Ro_{\rm critical}$=2.0.}
\label{fig:omega_fig}
\end{figure}

In Fig. \ref{fig:omega_fig} we highlight the rotational evolution of our sample of selected solar twins (red triangles), open clusters (black squares) and the Sun (in black, represented by the $\odot$ symbol). Stars with measured \Prot are represented by the black circles. We denoted as black crosses the centroid of each one of the 2 Gyr age-bin based on its respective age-\Prot average and dispersion (see Sec. \ref{sec:rotation_sample}). The solar twin sample is composed of those stars with \Prot errors within 1$\sigma$ from its respective age-\Prot cluster centroid. All selected stars are fully consistent with expected dispersion due to intrinsic measurement errors ($\sim$10\%). The only outlier is the 6 Gyr-old KIC 10130039 which deviates from the expected \Prot distribution by more than 50\%. Thanks to our extensive radial-velocity monitoring together with detailed mapping of chemical anomalies and activity levels \citep{santos17,spina17,lorenzo18}, we found that the stars placed considerably below the lower limit of rotation rate for a given age are, in fact, spectroscopic binaries. Two illustrative examples from \citet{santos16} are the 4.0 and 7.2 Gyr-old SB HIP 19911 and 67620 which rotate at 4.1 km/s and 2.7 km/s level (\Psini$\sim$12 and 20 days), respectively. We found that these stars are not likely to be representative of the rotational sample with a significance higher than 3$\sigma$. 
%@arxiver{FIG3_PROT_RO.pdf}
To test the hypothesis of weakened magnetic braking at a given Rossby number threshold \citep{vansaders16}, we constructed rotational evolution tracks adopting the well-known modified Kawaler wind-law \citep[assuming N=1.5,][]{kawaler88,krishnamurthi97} and the updated \textrm{YaPSI} grid of stellar tracks provided by \citet{spada17} for different masses and metallicities. The angular momentum equation is solved assuming a negligible moment of inertia change ($dI/I \rightarrow 0$) along the main-sequence: 
\begin{equation}\label{eq:kw}
\frac{dJ}{dt} = -K_w\Omega^3\Bigg(\frac{R}{R_\odot}\Bigg)^{0.5}\Bigg(\frac{M}{M_\odot}\Bigg)^{-0.5}, \qquad\qquad \rm{for}\quad \Omega>\Omega_{sat}  
\end{equation}
The constant $K_w$ is fine-tuned to match the unsaturated rotational evolution model to the solar properties (\Prot=25.4 days at 4.57 Gyr), considering 1.0 M/M$_\odot$ and solar metallicity track. The saturated value of rotation rate ($\Omega_{sat}$) is scaled by the solar convective turnover time following \citet{krishnamurthi97}. The rotational evolution model is modified whenever the star approaches into the turn-off (${\rm TO}$) region or a given critical Rossby number. After this stage, there is a dominance of structural changes over the magnetic braking terms in the angular momentum evolution (i.e. $dJ/dt\approx0$). Thus, we fix the angular momentum ($J_{\rm TO/crit}$), leaving only the moment of inertia to vary towards the main-sequence turn-off \citep{vansaders13}. For the Sun, our turn-off threshold is at $Ro_{\rm TO}\gtrsim$2.6 (or $\sim$8 Gyr).% which is considerably after the midlife $Ro_{\rm critical}$ of 2 recently adopted in the literature.

\begin{figure}
	\centering
	\begin{minipage}[t]{0.85 \linewidth}
		\centering
		\resizebox{\hsize}{!}{\includegraphics{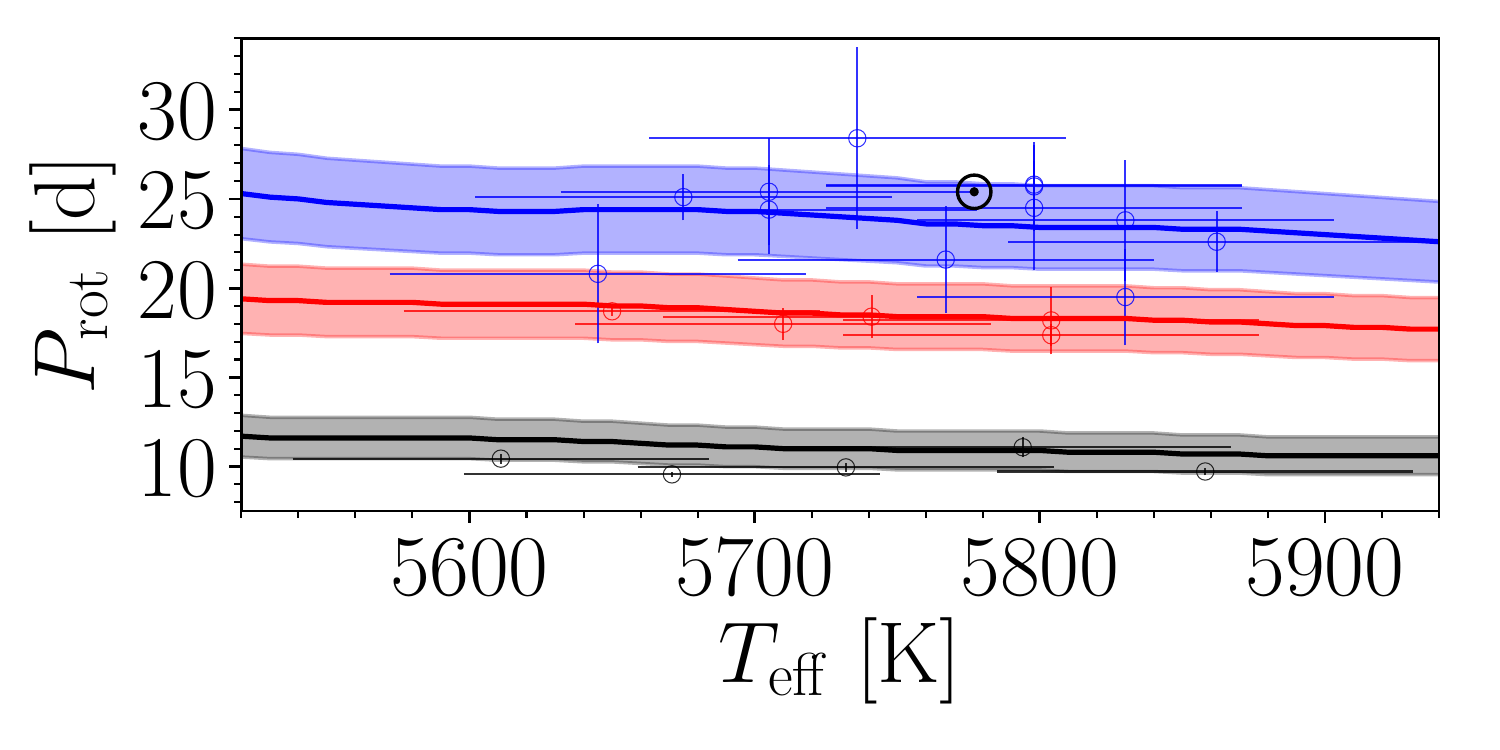}}
	\end{minipage}
	
	\caption{Consistency check of our models using OC data. Blue, red and black lines are the rotational tracks for M67, NGC6819 and NGC6811 with ages of 3.9, 2.4 and 0.85 Gyr, respectively. Each OC is represented by its correspondent color. The Sun is represented by $\odot$.}
	\label{fig:age_mass_oc}
\end{figure}

Predicted \Prot were derived for 22 Sun-like members of the OCs NGC6811, NGC6819 and M67 using Eq. \ref{eq:kw}. Effective temperatures were adopted as mass proxy using $(B-V)$ calibration by \citet{casagrande10} and literature spectroscopic \feh of each OC \citep{leebrown15,liu16b,netopil16}. In other words, the rotational tracks were built as function of age, $T_{\rm eff}$ and ${\rm [Fe/H]}$. In Fig. \ref{fig:age_mass_oc}, we test the consistency of our predicted \Prot as a function of $T_{\rm eff}$, fixing the age of each OC. The 1$\sigma$ confidence bands set by 10\% of \Prot uncertainty are represented by the shaded regions around each OC age-\Teff diagram prediction. Visually, our predictions are in good agreement with the existent \Prot data. For NGC6811, we show a rotational track for an age of 0.85 Gyr which is 15\% younger than the canonical age of 1 Gyr. 

We show in Fig. \ref{fig:omega_fig} the $Ro_{\rm TO}\sim$2.6 (left panel) and $Ro_{\rm crit}\sim$2 (right panel) scenarios for the rotational evolution of the Sun. Visually, both approaches seem to be in reasonable agreement. The magnetic weakened braking scenario favors the faster rotators placed in the lower boundary of the \Prot distribution. On the other hand, the model with smooth rotational evolution follows more closely the average trend observed, especially towards the oldest stars.

We quantified the suitability of both approaches by computing for each star the \Prot probability density function that depends on the age, mass, \feh for a given $Ro_{\rm crit}$. The errors are always assumed to follow Gaussian distributions. The adopted theoretical \Prot is represented by its median ($P_{\rm rot}^P$) and 16$-$84\% percentiles based on \Prot cumulative distribution function ($\sigma_{P_{\rm rot}}^P$). For instance, we derived $Ro_{\odot}$=2.13 ($P_{\rm rot, \odot}^P$=25.3 days). We calculated the Bayesian Information Criterion ($BIC$). $BIC$ accounts the trade-off between the fitting quality ($\hat{L}$), number of fitting parameters ($k$) of a given model (M) and the sample size (N): $BIC({\rm M}) = -2\ln\hat{L}(M)+k\ln{N}$, where $\hat{L}(M)$ is the product of likelihood of each data point with the composite errors computed through quadratic propagation of the individual measured and predicted \Prot errors. The $BIC$ difference ($\Delta BIC({\rm M2},{\rm M1}) \equiv BIC_{\rm M2} - BIC_{\rm M1}$) derived from 2 different models indicates which one is more likely. Defining M1 as the smooth rotational evolution model (Fig. \ref{fig:omega_fig}, left panel), we calculated $\Delta$BIC(M2,M1) where M2 stands for models with progressively larger $Ro_{\rm crit}$ ranging from 1.5 up to the subgiant branch, where both assumptions converge into the same \Prot solution ($Ro_{\rm crit}\rightarrow Ro_{\rm TO}$). The OC stars were used in this work as a consistency check of our models so our statistical tests are only based on field stars. 

To find the optimal $Ro_{\rm crit}$ for our sample, all the possible uncertainties were considered (10\% error related to SDR, model errors due to stellar parameters and measured \Prot errors). The best fit in terms of $Ro_{\rm crit}$ is $2.6_{-0.1}^{+\infty}$ with the corresponding 1$\sigma$ lower age limit of $t_{\rm crit}\gtrsim8.8$, $\gtrsim6.5$, $\gtrsim4.2$, and $\gtrsim2.5$ Gyr for 0.95, 1.00, 1.05 and 1.10 solar mass and metallicity stars, respectively. On the other hand, at 95\% confidence level, our result marginally approaches to the solar properties with the $Ro_{\rm crit}$ ranging from 2.3 to $Ro_{\rm TO}$ ($t_{\rm crit}\gtrsim$5.3 Gyr, for a solar mass/metallicity star). In all cases, the $\Delta$BIC analysis indicates values greater than +2. For $Ro_{\rm crit,\odot}$, we found $\Delta$BIC=+9.2 indicating a strong evidence favoring the smooth rotation evolution model, at least until the solar rotational level ($\gtrsim$25 days). For a solar mass star, if the magnetic weakened braking scenario is taking place at these $Ro$ thresholds, we should only detect unusually fast rotators (i.e. stars that depart the gyrochronology relations, considering the measurement errors) at ages considerably older than the Sun (say $\gtrsim$6 Gyr). Unfortunately, there is a lack of \Prot detections in this age range.

Given the statistical difficulties to disentangle the two scenarios, we tend to favor the simplest assumption of the smooth rotational evolution. Other possibility is that maybe the magnetic transition might occur, if it happens at all, at later evolutionary stages than it was hypothesized before \citep{vansaders16}. All in all, we conclude with a marginal level of confidence that, considering the available data of solar twins, no indisputable indication emerged about the weakened magnetic braking scenario. On the other hand, we are aware that this phenomenon might be correlated to other manifestations such as drastic changes in stellar cycle morphology and also in stellar differential rotation profile, as some \kepler\ data suggests. For a comprehensive discussion of these possibilities, see \citet{metcalfe17}. Even though, we stress that more data of similar stars is still needed to clarify this issue and firmly establish at what level should we trust on rotation-based ages. 

\section{Conclusions}\label{sec:conclusions}
The goal of this paper is to test different rotational evolution scenarios using a selected sample of solar twins characterized with the HARPS, HIRES and GRACES spectrographs. Stellar ages and other evolutionary parameters were estimated through HR diagram analysis with the help of new \gaia\ DR2 G band photometry and parallaxes and precise atmospheric parameters. Measured \Prot of 14 solar twins were collected from the literature and/or estimated in this work through \ion{Ca}{II} H \& K activity time-series. To trace the rotational evolution of solar twins, we build a grid of rotational evolutionary tracks based on modified Kawaler wind law and structural models. We compute these rotation tracks for a large range of critical Rossby number ($Ro_{\rm crit}$) to account for the magnetic weakened braking phenomena observed by \citet{vansaders16}. We found a marginal statistical evidence favoring the smooth rotation evolution. In the light of magnetic weakened braking scenario, the lower limit of critical Rossby number would be $Ro_{\rm crit}\gtrsim2.3$ (at 95\% confidence level) which intercepts an age range somewhat older than the Sun and the end of the main-sequence. This result highlights the difficulty to statistically discern both scenarios with the current sample of solar twins. Therefore, it is desirable that other works also approach this issue by determining new \Prot of old solar twin stars to clarify the past and the future of the solar rotational evolution.

\section*{Acknowledgements}
We  would  like  to  acknowledge  the  anonymous  referee, whose  comments  have  unquestionably  led  to  an  improved  paper. DLO is grateful to the Brazilian workers and taxpayers, whose effort provided in the past years the stability for young scientists to independently develop their scientific ideas. DLO and JM thanks support from FAPESP (2016/20667-8; 2018/04055-8). LAdS acknowledges the financial support from the European Research Council (ERC) under the European Union’s Horizon 2020 research and innovation program ({\sc Four Aces}; grant 724427). MA acknowledges funding from the Australian Research Council (grant DP150100250).% This research made use of the SIMBAD data base, operated at CDS, Strasbourg, France. 
%
%%%%%%%%%%%%%%%%%%%%%%%%%%%%%%%%%%%%%%%%%%%%%%%%%%



\bibliographystyle{mnras}
\bibliography{mnras_template} % if your bibtex file is called example.bib

%\begin{thebibliography}{99}

%\end{thebibliography}

%%%%%%%%%%%%%%%%%%%%%%%%%%%%%%%%%%%%%%%%%%%%%%%%%%

%%%%%%%%%%%%%%%%% APPENDICES %%%%%%%%%%%%%%%%%%%%%


%%%%%%%%%%%%%%%%%%%%%%%%%%%%%%%%%%%%%%%%%%%%%%%%%%

% Don't change these lines
\bsp	% typesetting comment
\label{lastpage}
\end{document}